\documentclass{article}

\usepackage{graphicx}
\usepackage{authblk}

\title{Proposal of a new quantum  annealing schedule for studying quantum annealing of transverse field Ising models}

\author{
Chiaki Yamaguchi\thanks{Email:info@to-qc.com}
}

\affil{
Tokyo Quantum Computing, Haramura 17217-1689, Suwagun, Nagano 391-0115, Japan
}

\date{}

\begin{document}
\maketitle

\begin{abstract}
Recently, Heim, R\o nnow, Isakov and Troyer [{\it Science} \textbf{348} (2015) 215]
 have reported that
 Monte Carlo simulations for the Ising spin glass model on the square lattice
 in the physically relevant continuous-imaginary-time limit
 do not show superiority of quantum annealing (QA) using transverse field
 against classical annealing (CA).
 Although the QA schedule that they had used has been using conventionally,
 however the QA schedule mathematically has no guarantee
 that the used schedule is the best QA schedule for performance of optimization.
 We propose a new QA schedule
  for studying transverse-field-based quantum versus classical annealing
 of the Ising model.
 The present QA schedule utilizes a smallest effective transverse field derived in this article.
 This QA schedule is made for
 the comparison between the system with no transverse field
 and the system with the smallest effective transverse field.
 As a case study,
 we study QA of the Ising spin glass model on the square lattice
 at low but finite temperature.
  A Monte Carlo algorithm
 using the physically relevant continuous-imaginary-time limit is performed.
 As the simulation results, we show superiority of QA against CA
 when the annealing time is sufficiently spent.
\end{abstract}


\section{Introduction} \label{sec:introduction}

 The quantum annealing (QA) is an optimization method utilizing quantum effect\cite{RCC, FGSSD, KN, ST, DC, TTC}. 
 By using QA, the solutions of given optimization problems are approximately or exactly obtained.
 QA is also related to a quantum computation by adiabatic evolution\cite{FGGS, FGGLLP}.

QA is proposed as an alternative of the classical annealing (CA)\cite{GG}.
In CA, the temperature is operated from high to low
 in order to utilize thermal fluctuations.
 On the other hand, in QA,
 external fields (and/or exchange interactions) generating quantum effect
 are operated from large to zero in order to utilize quantum fluctuations.
 These annealings help to make the systems escape
 from local minima of the free energy, and
 the systems reach low-energy states effectively.

The understandings of QA
 are recently becoming significant in relation to the D-Wave chip\cite{JAGetc}.
This chip is designed to perform QA for solving optimization problems.
 The occurrence of quantum properties for this chip is shown\cite{JAGetc}.

 We simulate QA of real-physical systems in this article,
 and we do not simulate QA as a classical optimization method in this article.
 Recently, it is reported in Ref.~\cite{HRIT} that
 Monte Carlo simulations for the Ising spin glass model on the square lattice
 in the physically relevant continuous-imaginary-time limit
 do not show superiority of QA using transverse field against CA.
 In order to study physically relevant systems,
 we use a continuous-imaginary-time quantum Monte Carlo algorithm 
 in this article.
 In addition, we investigate the Ising spin glass model on the square lattice.

 Although the QA schedule used in Ref.~\cite{HRIT} has been using conventionally,
 however the QA schedule mathematically has no guarantee
 that the used schedule is the best QA schedule for performance of optimization.
 We propose a new QA schedule
 for studying transverse-field-based quantum versus classical annealing
 of the Ising model.
 This QA schedule utilizes a smallest effective transverse field derived in this article.
 We derive this field based on a discussion of percolation of spin correlation
 per spin along the continuous-imaginary-time direction.
 The present QA schedule is made for the comparison between the system with no transverse field  and the system with the smallest effective transverse field.
 By this comparison, the superiority of QA (or CA) can be made clear.

We apply a quantum Monte Carlo algorithm \cite{NI} proposed by Nakamura and Ito.
 This algorithm uses the physically relevant continuous-imaginary-time limit.
 This algorithm is a Glauber-dynamics-like algorithm (a heat-bath-like algorithm) which performs single-spin flips in the real space and cluster flips in the imaginary-time space.
 The Glauber dynamics is suited to the study of the dynamical features of physical systems\cite{O, D}.
 By applying this algorithm,
 observing a true relaxation of the original system is expected\cite{NI}.
 The mathematical form of this algorithm is directly related to the derivation of a smallest effective transverse field derived in this article,
 thus we use this algorithm in this article

 We compare three annealings:
 the present one, a conventional quantum one (used for a quantum computation by adiabatic evolution) and a classical one.
 The simulation results for the annealings are shown in this article.

 This article is organized as follows.
 The quantum annealing and the spin glass model are explained
 in \S\ref{sec:Model}.
 The derivation of
  a quantum Monte Carlo algorithm by Nakamura and Ito \cite{NI}
 is described in \S\ref{sec:QuantumMonteCarloMethod1}.
 The way of applying this algorithm is written
 in \S\ref{sec:QuantumMonteCarloMethod2}.
 The differences between this algorithm and the related algorithms \cite{RK, MSN} are mentioned
 in \S\ref{sec:QuantumMonteCarloMethod2}.
 Quantities that we treat in this article are described
 in \S\ref{sec:Quantities}.
 A smallest effective transverse field is derived in \S\ref{sec:AnnealingSchedule}.
 A new QA schedule is proposed in \S\ref{sec:AnnealingSchedule}.
 Simulation results are shown in \S\ref{sec:SimulationResults}.
 Concluding remarks are in \S\ref{sec:ConcludingRemarks}.

\section{The quantum annealing and the spin glass model} \label{sec:Model}

In terms of theoretical physics,
 the quantum annealing can be to investigate
 a time dependent Hamiltonian ${\cal H} (t)$ which has
 quantum effect terms.
 Usually, the Hamiltonian ${\cal H} (t)$ can be written as
\begin{equation}
 {\cal H} (t) = J (t) {\cal H}_P + \gamma (t) {\cal H}_Q \, , \label{eq:HJg}
\end{equation}
 where $t$ is the time from $t = 0$ to $t = T_I$.
 ${\cal H}_P$ is the problem Hamiltonian that an optimization problem is written.
 ${\cal H}_Q$ is a quantum effect Hamiltonian for generating quantum effect.
 When $T_I$ is a small number, this annealing
 corresponds to a fast annealing,
 and, when $T_I$ is a large number,
 this annealing corresponds to a slow annealing.
 $J (t)$ is an increasing function for $t$, or $J (t)$ is a constant.
 $J (t)$ increases from $0$ to $1$ for example.
 $\gamma (t)$ is a decreasing function for $t$.
 $\gamma (t)$ decreases from a large value to zero for example, or
 $\gamma (t)$ decreases from $1$ to $0$ for example.
 A method using a pulse of the transverse field is also proposed\cite{MSN}.

 When $\gamma (t)$ gives zero from $t = 0$ to $t = T_I$
 while the value of $J (t)$ (or the temperature) changes,
 this annealing is a classical annealing.
 When there is a time that $\gamma (t)$ gives non-zero,
 this annealing is a quantum annealing.
 
The problem Hamiltonian ${\cal H}_P$ is written as the form of
 the Ising spin glass model called the Edwards-Anderson model.
The Hamiltonian ${\cal H}_P$ is given by  \cite{EA, MPV, N, KR}
\begin{equation}
 {\cal H}_P = - \sum_{< i j >} \tau_{i j} S_i S_j  \, , \label{eq:H1}
\end{equation}
 where $< i j >$ denotes nearest-neighbor pairs, $S_i$ is
 a state of a spin at site $i$, $S_i = \pm 1$,
 and this spin is called the Ising spin.
 The value of $\tau_{i j}$ is given by the problem that is asked to solve.
 In the spin glass model, the probability of giving the value of $\tau_{i j}$ is 
\begin{equation}
 P ( \tau_{i j} ) = \frac{1}{2} \delta_{\tau_{i j}, - 1} + \frac{1}{2} \delta_{\tau_{i j}, 1}
 \label{eq:Ptau} 
\end{equation}
 for  example.
 Here, $\delta$ is the Kronecker delta.
 We use the value of $\tau_{i j}$ obtained by using Eq.~(\ref{eq:Ptau})
 in this article.
The Ising spin glass model using $P ( \tau_{i j} )$
 is called the $\pm J$ Ising spin glass model.
 We investigate the $\pm J$ Ising spin glass model
 on the square lattice with periodic boundary conditions in this article.

The quantum effect Hamiltonian ${\cal H}_{Q}$,
 which is investigated in this article,
 is given by \cite{RCC, KN, ST, DC}
\begin{equation}
 {\cal H}_{Q} = - \sum_{i = 1}^N \sigma^x_i \, , \label{eq:HQ}
\end{equation}
 where $\sigma^k_i$ is the k component of the Pauli matrix at site i.
 $N$ is the number of sites (spins) in the whole system.
 The Pauli matrices are
\begin{eqnarray}
  \sigma^x =
  \left(
   \begin{array}{cc}
    0 & 1\\
    1 & 0
   \end{array}
  \right),
  \quad
  \sigma^y =
  \left(
   \begin{array}{cc}
   0 & -i\\
   i & 0
   \end{array}
  \right)
  \quad {\rm and} \quad
  \sigma^z =
  \left(
   \begin{array}{cc}
   1 & 0\\
   0 & -1
   \end{array}
  \right) \, .
\end{eqnarray}
 When the measurement axis is chosen to be $z$,
 the external field by $\sigma^x$ is called the transverse field.
 An Ising spin under the transverse field corresponds to a quantum bit
 in this study.

 By using Eqs.(\ref{eq:HJg}), (\ref{eq:H1}) and (\ref{eq:HQ}),
 the Hamiltonian ${\cal H} (t)$ is also written as
\begin{equation}
 {\cal H} (t) = - J (t) \sum_{< i j >} \tau_{i j} \sigma^z_i \sigma^z_j
 - \gamma (t) \sum_{i = 1}^N \sigma^x_i 
 \,  . \label{eq:H2}
\end{equation}
 We study this Hamiltonian ${\cal H} (t)$.

\section{The derivation of a continuous-imaginary-time quantum Monte Carlo algorithm} \label{sec:QuantumMonteCarloMethod1}

 The derivation of
  a quantum Monte Carlo algorithm by Nakamura and Ito \cite{NI} is described in this section,
 since the mathematical form of this algorithm is directly related to
 the derivation of a smallest effective transverse field derived in this article.
 The way of applying this algorithm is described in
 \S\ref{sec:QuantumMonteCarloMethod2}.

The Glauber dynamics
is suited to the study of the dynamical features of physical systems\cite{O, D}.
 By applying the Glauber dynamics,
 discussions for dynamical features as a physical system are possible\cite{O, D}.
 The Glauber dynamics is a heat-bath algorithm for the Ising model.
 We use a Glauber-dynamics-like algorithm (a heat-bath-like algorithm) which
 performs
 single-spin flips in the real space and cluster flips in the imaginary-time space.
 For the real-space direction, this algorithm performs single-spin flips like
 as the Glauber dynamics does.
 For the imaginary-time direction,
 the continuous-imaginary-time limit is applied.

 We use the Suzuki-Trotter decomposition given by \cite{S}
\begin{equation}
 e^{- \beta ({\cal H}_A + {\cal H}_B + {\cal H}_C )}
 = \lim_{\Delta \tilde{t} \to 0} ( e^{- {\cal H}_A \Delta \tilde{t} }
 e^{- {\cal H}_B \Delta \tilde{t} } e^{- {\cal H}_C \Delta \tilde{t} } 
 )^{\frac{\beta}{\Delta \tilde{t}}} \, ,
\end{equation}
 where $\Delta \tilde{t} \equiv \frac{\beta}{M}$,
 $M$ is called the Trotter number,
 and $\tilde{t}$ is called the imaginary time.
 $\beta$ is the inverse temperature,
 $\beta = 1 / ( k_B T_E )$, $T_E$ is the temperature,
 and $k_B$ is the Boltzmann constant.
 The physically relevant continuous-imaginary-time limit
 mentioned in Ref.~\cite{HRIT} corresponds to the limit of $\Delta \tilde{t} \to 0$ in this article.
 By this decomposition,
 the $d$-dimensional quantum system is treated as
 the $d + 1$-dimensional classical system.
 A continuous-imaginary-time quantum Monte Carlo algorithm is used in this study.
 This means that
 the system for $\Delta \tilde{t} \to 0$ is directly investigated.
 This study is not affected by the error of the Suzuki-Trotter decomposition.
 There is a relation\cite{S}:
\begin{equation}
 \langle \sigma^{z}_x | e^{\beta \gamma (t) \sigma^x} | \sigma^{z}_y \rangle
 = \frac{1}{2} ( e^{\beta \gamma (t)} +  \sigma^{z}_x \, \sigma^{z}_y e^{ - \beta \gamma (t)} ) \, ,
\end{equation}
 where $\sigma^{z}_x, \sigma^{z}_y = \pm 1$.
 Thus, the partition function $Z (t)$ is written as
 \cite{S}
\begin{eqnarray}
 & & Z (t) = \lim_{\Delta \tilde{t} \to 0}  \sum_{ \{ \sigma^z_{i k} \} }
 \exp \biggl\{ J (t) \Delta \tilde{t} \sum_{k \in \rho} \sum_{ i j \in \rho}
 \tau_{i j} \sigma^z_{i k} \sigma^z_{j k} 
  \nonumber \\ & & 
 + \sum_{k l \in \rho} \sum_{i \in \rho} \log \frac{
 \exp ( \gamma (t)  \Delta \tilde{t} )
 + \sigma^z_{i k} \sigma^z_{i l}
 \exp ( - \gamma (t)  \Delta \tilde{t} )}{2} \biggr\} \, ,
 \label{eq:Z}
\end{eqnarray}
 where $\rho$ is the set of the coordinates of the weights
 for the decomposed partition function in the $d + 1$-dimensional classical system.
 The coordinates of $\rho$ is somewhat complicated\cite{S},
 but there is no need to understand the coordinates, because of taking
 the continuous-imaginary-time limit.
 By performing $\Delta \tilde{t} \to 0$,
 the imaginary time becomes a continuous space.
 In Eq.~(\ref{eq:Z}), the term for $\sigma^z_{i k} \sigma^z_{j k}$
 is a term for the interaction of real-space direction,
 and the term for $\sigma^z_{i k} \sigma^z_{i l}$ is
 a term for the interaction of imaginary-time direction.
 The imaginary-time direction has periodic boundary conditions.
 The size of the imaginary-time direction is
 the same with the inverse temperature $\beta$.
 Eq.~(\ref{eq:Z}) is also written for $\Delta \tilde{t} \to 0$ as
\begin{eqnarray}
 Z (t) &=& \lim_{\Delta \tilde{t} \to 0}  \sum_{ \{ \sigma^z_{i k} \} }
 \exp \biggl[ J (t) \Delta \tilde{t} \sum_{k \in \rho} \sum_{ i j \in \rho}
 \tau_{i j} \sigma^z_{i k} \sigma^z_{j k} \nonumber \\
 & & + \log ( \gamma (t)  \Delta \tilde{t} )
 \sum_{k l \in \rho} \sum_{i \in \rho} \frac{
 1 - \sigma^z_{i k} \sigma^z_{i l} }{2} \biggr] \, .
 \label{eq:Z2}
\end{eqnarray}

 In order to derive this algorithm, we derive weights for graph representation.
 The framework of deriving weights for graph representation is described in Ref.~\cite{KG}.
 We define the weight of two spins along imaginary-time direction
 as $w (\sigma^z_{i k}, \sigma^z_{i l})$. 
 $w (\sigma^z_{i k}, \sigma^z_{i l})$ is given by
\begin{equation}
w (\sigma^z_{i k}, \sigma^z_{i l})
 = \exp \biggr[ \log ( \gamma (t)  \Delta \tilde{t} )
 \, \frac{ 1 - \sigma^z_{i k} \sigma^z_{i l} }{2} \biggr] \, .
 \label{eq:a-1}
\end{equation}
We define the weight for $\sigma^z_{i k} \sigma^z_{i l} = 1$
 as $w_{\rm para}$. By using Eq.~(\ref{eq:a-1}), we obtain
\begin{equation}
 w_{\rm para} = 1 \, . \label{eq:a-2}
\end{equation}
We define the weight for $\sigma^z_{i k} \sigma^z_{i l} = - 1$
 as $w_{\rm anti}$. By using Eq.~(\ref{eq:a-1}), we obtain
\begin{equation}
 w_{\rm anti} = \gamma (t) \Delta \tilde{t} \, . \label{eq:a-3}
\end{equation}
 We define the weight of graph for connecting two spins as $w (g_{\rm conn})$.
 We define the weight of graph for cutting two spins
 as $w (g_{\rm cut})$.
 We are able to write 
\begin{equation}
 w_{\rm para} = w (g_{\rm conn}) + w (g_{\rm cut}) \, , \label{eq:a-5}
\end{equation}
 and
\begin{equation}
 w_{\rm anti} = w (g_{\rm cut})  \, . \label{eq:a-6} 
\end{equation}
 By using Eqs.~(\ref{eq:a-2}) - (\ref{eq:a-6}),
 we obtain
\begin{equation}
 w (g_{\rm conn}) = 1- \gamma (t) \Delta \tilde{t} \, , \label{eq:a-7}
\end{equation}
 and
\begin{equation}
 w (g_{\rm cut}) = \gamma (t) \Delta \tilde{t}  \, . \label{eq:a-8}
\end{equation}
We define the probability of
 cutting two spins for $\sigma^z_{i k} \sigma^z_{i l} = 1$ as
 $P_{\rm para} (g_{\rm cut})$.
 We define the probability of cutting two spins for $\sigma^z_{i k} \sigma^z_{i l} = - 1$
 as $P_{\rm anti} (g_{\rm cut})$.
 We are able to write
\begin{equation}
 P_{\rm para} (g_{\rm cut}) =
 \frac{w (g_{\rm cut})}{w (g_{\rm conn}) + w (g_{\rm cut})} =
 \gamma (t) \Delta \tilde{t} \, , \label{eq:a-9} 
\end{equation}
 and
\begin{equation}
 P_{\rm anti} (g_{\rm cut})
 = \frac{w (g_{\rm cut})}{w (g_{\rm cut})} =
 1 \, . \label{eq:a-10}
\end{equation}

 We calculate the probability of giving the number of cuts per spin
 for $P_{\rm para} (g_{\rm cut})$ of Eq.~(\ref{eq:a-9})
 in the continuous-imaginary-time limit.
 We define $n_{\rm add}$ as the number of cuts per spin
 for parallel spins $\sigma \, \sigma' = 1$ along the imaginary-time direction.
 By using Eq.~(\ref{eq:a-9}), the probability $P (n_{\rm add})$ of giving $n_{\rm add}$
 is obtained as
\begin{eqnarray}
 P (n_{\rm add} ) &=&  \lim_{\Delta \tilde{t} \to 0}  
 \left( \begin{array}{c} \frac{\beta}{\Delta \tilde{t}} - n_{\rm kink} \\ n_{\rm add} \end{array} \right)
 ( 1 - P_{\rm para}  )^{\frac{\beta}{\Delta \tilde{t}} - n_{\rm kink} - n_{\rm add} } ( P_{\rm para}  )^{n_{\rm add}}
 \nonumber \\
 &=& \frac{1}{n_{\rm add} !} ( \gamma (t)  \beta )^{n_{\rm add}}
 \exp ( - \gamma (t)  \beta ) \, , \label{eq:a-11}
\end{eqnarray}
 where $\left( \begin{array}{c} x \\ y \end{array} \right) \equiv \frac{x !}{y ! (x - y) !}$,
 and $n_{\rm kink}$ is the number of kinks per spin.
 The kink means the position of antiparallel spins
 $\sigma \, \sigma' = -1$ along the imaginary-time direction.
 The position $\tilde{t}_{\rm add}$ of the cut for parallel spins $\sigma \, \sigma' = 1$
 along the imaginary-time direction is obtained by
\begin{equation}
 \tilde{t}_{\rm add} = \beta R \, , \label{eq:a-12}
\end{equation}
 where $R$ is a pseudorandom number for $0 \le R < 1$.
 By using Eq.~(\ref{eq:a-10}),
 cuts are added for antiparallel spins $\sigma \, \sigma' = - 1$
 along the imaginary-time direction with probability one.
 Thus the number of cuts per spin for antiparallel spins
 along the imaginary-time direction is $n_{\rm kink}$.
 Therefore, the number of all cuts per spin along the imaginary-time direction, $n_{\rm cut}$,
 is given by 
\begin{equation}
 n_{\rm cut} = n_{\rm add} + n_{\rm kink} \, . \label{eq:a-13}
\end{equation}

 This algorithm performs single-spin flips in the real space.
 The transition probability $P ( E \to E ' )$ for the energy transition $E \to E '$ is given by 
\begin{equation}
 P ( E \to E ' )
 = \frac{e^{ - \beta E '}  }
 {e^{- \beta  E} + e^{ - \beta E '}  } \, . \label{eq:a-14}
\end{equation}
 Cuts are added along the imaginary-time direction,
 and the spin correlations along the imaginary-time direction
 are cut at the positions of the cuts.
 Therefore, $P ( E \to E ' )$ is only calculated for real spaces.
 When there is no cut,
 this algorithm becomes the Glauber dynamics.

\section{The way of applying this Monte Carlo algorithm} \label{sec:QuantumMonteCarloMethod2}

 We describe the way of applying this algorithm.
 The derivation of this algorithm is in \S\ref{sec:QuantumMonteCarloMethod1}.
 When the transverse field is not imposed,
 this algorithm becomes the Glauber dynamics.

One Monte Carlo step (sweep) is as follows:
\begin{enumerate}
\item The number of cuts for parallel spins along the imaginary-time direction, $n_{\rm add}$, is calculated by using Eq.~(\ref{eq:a-11}) for each spin, and the positions of cuts for parallel spins along the imaginary-time direction, $\tilde{t}_{\rm add}$, are calculated $n_{\rm add}$ times by using Eq.~(\ref{eq:a-12}) for each spin. Moreover, cuts are placed on all kinks with probability one (Eq.~(\ref{eq:a-10})). Clusters are detected by all the cuts. The number of clusters, $N_C$, is counted up. When there is no cut, $N_C$ gives $N$ (the number of spins).
\item One of the detected clusters is randomly picked up, and the cluster is flipped with the probability of Eq.~(\ref{eq:a-14}). The trials of flipping clusters are performed $N_C$ times.
\item All the cuts are deleted.
\item Quantities are sampled if needing to sample the quantities.
\end{enumerate}
 One Monte Carlo step is a unit of time used in this study.
 This time is called the Monte Carlo time.
 We investigate physical systems under this time evolution.

\begin{figure}[t]
\begin{center}
\includegraphics[width=0.68\linewidth]{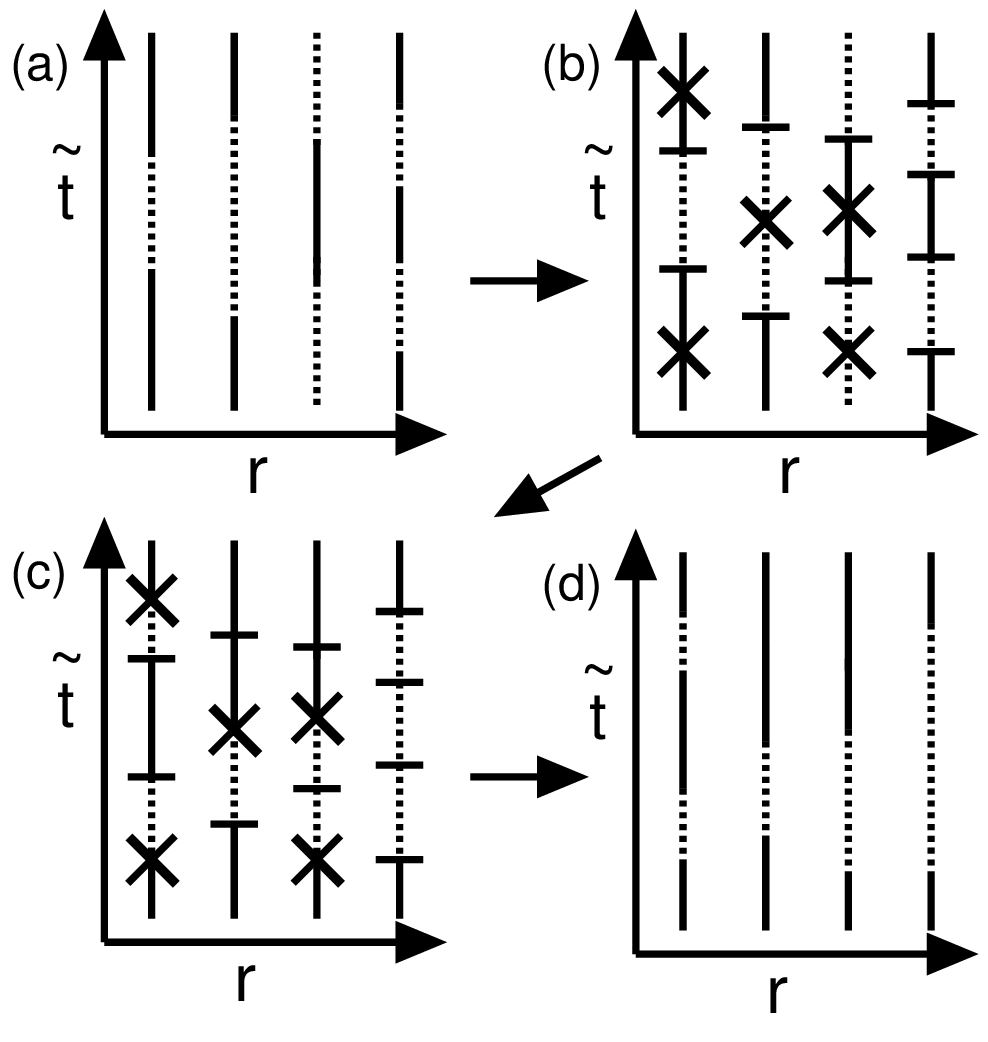}
\end{center}
\caption{
 A sketch of one Monte Carlo step.
 Full lines correspond to worldline segments with spin up,
 and broken lines correspond to worldline segments with spin down.
 The horizontal axis indicates the real space, and
 the longitudinal axis indicates the imaginary time.
 (a) Spin states are shown.
 (b) Cuts are added.
 Cuts for kinks are represented as bars,
 and cuts for parallel spins along the imaginary-time direction
 are represented as crosses.
 (c) Cluster flips are performed.
 (d) Cuts are deleted, and quantities are sampled. 
\label{fig:figure-1}
}
\end{figure}
 Fig.~\ref{fig:figure-1} shows
 a sketch of one Monte Carlo step.
 Full lines correspond to worldline segments with spin up, and
 broken lines correspond to worldline segments with spin down.
 The horizontal axis indicates the real space, and
 the longitudinal axis indicates the imaginary time.
 A discrete space is assumed for the real-space direction, and
 a continuous space is assumed for the imaginary-time direction.
 The imaginary-time direction has periodic boundary conditions.
 (a) Spin states are shown.
 (b) Cuts are added.
 Cuts for kinks are represented as bars,
 and cuts for parallel spins along the imaginary-time direction
 are represented as crosses.
 Cuts for kinks are added with the probability one according to Eq.~(\ref{eq:a-10}).
 Cuts for parallel spins along the imaginary-time direction
 are added according to Eqs.(\ref{eq:a-11}) and (\ref{eq:a-12}).
 (c) Cluster flips are performed.
 One of clusters is randomly picked up, and 
 the cluster is flipped with the probability of Eq.~(\ref{eq:a-14}).
 Cluster flips are tried for the number of clusters.
 (d) Cuts are deleted, and quantities are sampled. 

\begin{figure}[t]
\begin{center}
\includegraphics[width=0.68\linewidth]{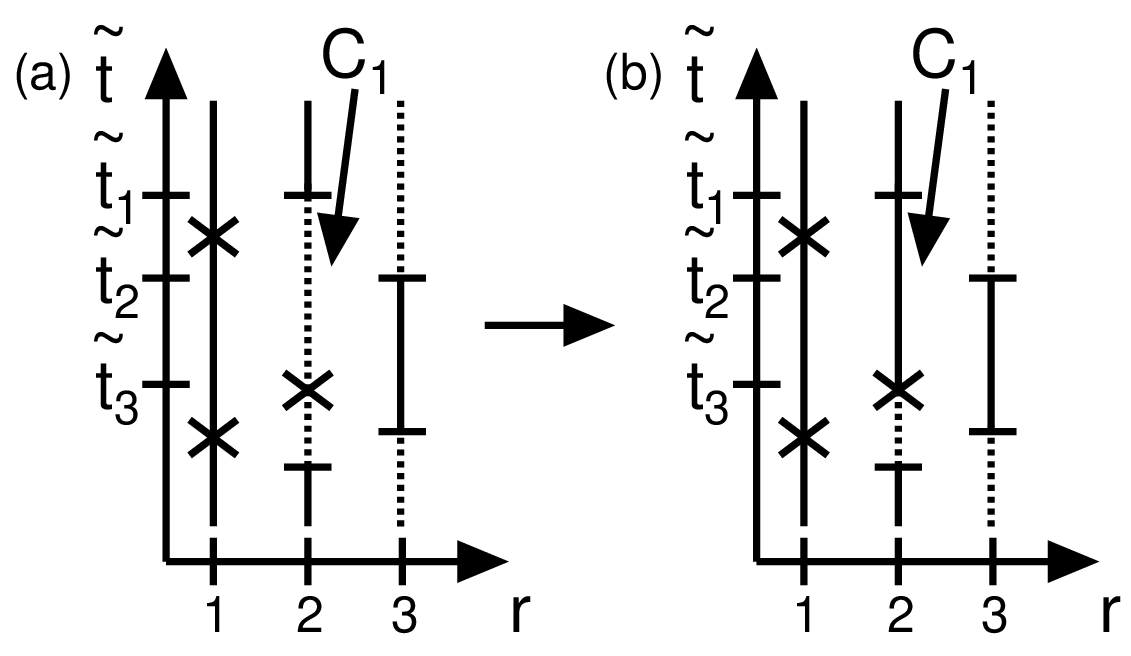}
\end{center}
\caption{
 A sketch of one cluster flip.
 The horizontal axis indicates the real space,
 and the longitudinal axis indicates the imaginary time.
 (a) Spin states are shown before a cluster $C_1$ is flipped.
 (b) Spin states are shown after the cluster $C_1$ was flipped.
\label{fig:figure-2}
}
\end{figure}
 Fig.~\ref{fig:figure-2} shows a sketch of one cluster flip.
 The horizontal axis indicates the real space,
 and the longitudinal axis indicates the imaginary time.
 (a) Spin states are shown before a cluster $C_1$ is flipped.
 (b) Spin states are shown after the cluster $C_1$ was flipped.
 If the transition probability $P ( E \to E ' )$ of Fig.~\ref{fig:figure-2} is calculated,
 $P ( E \to E ' )$ is obtained as
 $P ( E \to E ' ) = 1 / \{ 1 + \exp [ - 2 J (t) ( \tilde{t}_1 -  \tilde{t}_3 ) \tau_{1 2}
 + 2 J (t) (\tilde{t}_1 -  2 \tilde{t}_2 +  \tilde{t}_3 ) \tau_{2 3} ] \}$.

One simulation is as follows:
\begin{enumerate}
\item A problem (a set of  $\tau_{i j}$) is prepared.
\item Initial states of spins at $t = 0$ are made by using some Monte Carlo steps, and quantities are sampled for $t = 0$.
\item One Monte Carlo step is performed. $t \leftarrow t + 1$ is executed. Quantities are sampled for $t$.
\item When $t = T_I$, the simulation is closed.
\end{enumerate}
 By repeating the above procedures,
 many simulations are performed,
 and the average values of quantities are calculated. 

 We describe the differences between related algorithms.
The algorithm by Rieger and Kawashima is that
 for each site one generates new cuts in addition
to the old ones from the already existing segments
via a Poisson process with decay time $1 / \gamma$ along
 the imaginary-time direction\cite{RK}.
 This algorithm directly detects the correlation length along
 the imaginary-time direction.
 On the other hand, this algorithm directly treats the number of cuts
 for detecting the correlation lengths along the imaginary-time direction.
 This algorithm generates the number of cuts for each site,
 and this algorithm adds the cuts randomly.
 The algorithm by Morita, Suzuki and Nakamura is that the initial cluster state is made by giving the position of cuts by the Poisson process with the mean value $\beta \gamma$ \cite{MSN}.
 If the number of generated domain walls is odd, one of the domain walls is removed to obey the periodic boundary conditions along the imaginary-time direction\cite{MSN}.
 This algorithm does not judge whether the number of generated domain walls is odd or even.
 If this process for removing one of the domain walls was combined with this algorithm, this combined algorithm gives incorrect results.
 Therefore, this algorithm and the algorithm by Morita, Suzuki and Nakamura should be considered to be separated.

 The simulation results by these algorithms can be the same for all situations.
 The orders of calculation costs for this algorithm and the related algorithms
 can also be the same for the system sizes,
 thus we do not compare the algorithms in detail in this article.
 The aim of this article is to
 propose a new quantum annealing schedule and
 to investigate the efficiency of the schedule.
 The mathematical form of this Monte Carlo algorithm is
 directly related to the derivation of a smallest effective transverse field,
 and the schedule that we propose in this article utilizes the field.
 We derive the field in \S\ref{sec:AnnealingSchedule}.

\section{Quantities} \label{sec:Quantities}

We describe the quantities that we investigate in this article.

By using Eq.~(\ref{eq:Z2}), 
 the expectation value $[ \langle E \rangle_T ]_J$
 of the energy $E$ of the Hamiltonian ${\cal H} (t)$ (Eq.~(\ref{eq:H2})) is written as
\begin{eqnarray}
 & & [ \langle E \rangle_T ]_J = - \frac{\partial}{\partial \beta} [ \log Z ]_J \nonumber \\
 &=& \biggl[ \frac{1}{Z} \lim_{\Delta \tilde{t} \to 0} \sum_{ \{ \sigma^z_{i k} \} }
 \biggl( - \frac{J (t)}{M} \sum_{k \in \rho} \sum_{ i j \in \rho}
 \tau_{i j} \sigma^z_{i k} \sigma^z_{j k} 
 - \frac{1}{\beta} \sum_{k l \in \rho} \sum_{i \in \rho} \frac{
 1 \! - \! \sigma^z_{i k} \sigma^z_{i l} }{2} \biggr) \nonumber \\
 & & \! \! \! \times \exp \biggl[ J (t) \Delta \tilde{t} \sum_{k \in \rho} \sum_{ i j \in \rho}
 \tau_{i j} \sigma^z_{i k} \sigma^z_{j k} 
 + \log ( \gamma (t)  \Delta \tilde{t} )
 \sum_{k l \in \rho} \sum_{i \in \rho} \frac{
 1 \! - \! \sigma^z_{i k} \sigma^z_{i l} }{2} \biggr] \biggr]_J \, , \nonumber \\
 & & \label{eq:E} 
\end{eqnarray}
 where $\langle \, \rangle_T$ is the thermal average,
 and $[ \, ]_J$ is the average for $\tau_{i j}$.
 $[ \, ]_J$ is the average for the problems that are asked to solve,
 and, if the word for the spin glass model is used,
 $[ \, ]_J$ is the random configuration average.
 Therefore, in order to calculate the expectation value of the energy
 by the Monte Carlo simulation,
\begin{equation}
 E = - \frac{J (t)}{\beta} \sum_{< i j >} \int_0^\beta  \! d \tilde{t} \,
 \tau_{i j} \sigma^z_{i} ( \tilde{t} ) \sigma^z_{j} ( \tilde{t} ) 
 - \frac{1}{\beta} \, \sum_{i = 1}^N n_{\rm kink} (i) \, \label{eq:E2}
\end{equation}
 is sampled, 
 where $n_{\rm kink} (i)$ is the number of kinks at site $i$.
 By the way, the solution of a given optimization problem at site $i$ can be obtained as
\begin{equation}
 S_i^{\rm result} = \frac{\int_0^\beta \! d \tilde{t} \, \sigma^z_{i} ( \tilde{t} ) }
 {\big| \int_0^\beta \! d \tilde{t} \, \sigma^z_{i} ( \tilde{t} ) \big| }  \, . \label{eq:Sresult}
\end{equation}
 The aim of the quantum annealing is to solve optimization problems.
 Therefore, by using Eqs.(\ref{eq:H1}) and (\ref{eq:Sresult}),
 investigating an energy $E_P$ given by
\begin{equation}
 E_P = - \sum_{< i j >}
 \tau_{i j} \, \frac{\int_0^\beta \! d \tilde{t} \, \sigma^z_{i} ( \tilde{t} ) }
{\big| \int_0^\beta \! d \tilde{t} \, \sigma^z_{i} ( \tilde{t} ) \big| }
 \, \frac{\int_0^\beta \! d \tilde{t} \, \sigma^z_{j} ( \tilde{t} ) }
{\big| \int_0^\beta \! d \tilde{t} \, \sigma^z_{j} ( \tilde{t} ) \big| }
  \label{eq:E3} 
\end{equation}
 can be suited for this study,
 since, if a lot of kinks remains in the system after annealing,
 there is a possibility that $E$ of Eq.~(\ref{eq:E2}) gives an unexpected low value.
 In addition, there is a possibility that this unexpected low value
 gives a false impression for success of the quantum annealing.
 Thus we investigate the energy $E_P$ instead of the energy $E$ in this article.

 We define the exact ground-state energy of Eq.~(\ref{eq:H1}) as $E_G$,
 where the exact ground-state is the exact solution of the given problem.
 $E_G$ depends on $\tau_{i j}$.
 The residual energy $E_{\rm res}$ is given by
\begin{equation}
 E_{\rm res} = [ \langle E_P \rangle_T - E_G ]_J \, . \label{eq:Eres}
\end{equation}
 We investigate the residual energy $E_{\rm res}$ in this article. 

 The accuracy $P_{\rm exact}$ is given by
\begin{equation}
 P_{\rm exact} = [ \langle \delta_{E_P , E_G} \rangle_T ]_J  \, . \label{eq:Pexact}
\end{equation}
  We investigate the accuracy $P_{\rm exact}$ in this article.

\section{A new quantum annealing schedule} \label{sec:AnnealingSchedule}

 The mathematical form of this Monte Carlo algorithm
 derived in \S\ref{sec:QuantumMonteCarloMethod1} is
 directly related to the derivation of a smallest effective transverse field,
 and the quantum annealing schedule that we propose in this article utilizes the field.
 Firstly, we derive the field based on a discussion of percolation of spin correlation
 per spin along the continuous-imaginary-time direction.

 By cuts used in Eqs.(\ref{eq:a-10}) - (\ref{eq:a-12}),
 spin correlations are cut along the imaginary-time direction.
 We consider the fewest number of cuts per spin.
 Since the imaginary-time direction has periodic boundary conditions,
 by considering a percolation of spin correlations,
 the fewest number of cuts per spin,
 $n_{\rm fewest \, cut}$, is obtained as two, 
\begin{equation}
 n_{\rm fewest \, cut} = 2 \, . \label{eq:c-1}
\end{equation}
 By using Eqs.~(\ref{eq:a-13}) and (\ref{eq:c-1}),
 the fewest number of cuts per spin for parallel spins
 along the imaginary-time direction, $n_{\rm fewest \, add}$, is obtained as
\begin{equation}
 n_{\rm  fewest \, add} + n_{\rm kink} = 2 \, . \label{eq:c-1-2}
\end{equation}
 By using  Eq.(\ref{eq:a-11}),
 the average number of cuts per spin for parallel spins
 along the imaginary-time direction is obtained as
\begin{equation}
 \sum_{n_{\rm add} = 0}^{\infty} n_{\rm add} P ( n_{\rm add} ) = \gamma (t) \beta \, .
 \label{eq:c-2}
\end{equation}
 Therefore, when there is no kink ($n_{\rm kink} = 0$),
 by using Eqs.~(\ref{eq:c-1-2}) and (\ref{eq:c-2}),
 we obtain a smallest effective transverse field $\gamma_S$  as
\begin{equation}
 \gamma_S = \frac{2}{\beta} \, .  \label{eq:c-3}
\end{equation}
 When there are kinks,
 by using Eq.~(\ref{eq:c-1-2}), 
 $n_{\rm  fewest \, add}$ is obtained as
\begin{equation}
 n_{\rm  fewest \, add} < 2 \, , \label{eq:c-1-3}
\end{equation}
 since $n_{\rm kink} > 0$.
Therefore, when there are kinks,
 a smallest effective transverse field $\tilde{\gamma}_S$ is obtained as
 \begin{equation}
 \tilde{\gamma}_S < \gamma_S \, .  \label{eq:c-4}
\end{equation}
 If there is no kink, the transverse field $\gamma$ for $\gamma < \gamma_S$
 is not effective for generating quantum effect.
 $\gamma$ for $\gamma = \gamma_S$
 is effective for generating quantum effect regardless of the number of kinks.
 A smallest effect transverse field that does not depend on the number of kinks is
 desired in this study. 
 Therefore, we estimate that $\gamma_S$ is the smallest effective field
 that should be used in this study.

We propose a QA schedule
 for studying transverse-field-based quantum versus classical annealing
 of the Ising model by utilizing the field $\gamma_S$ as
\begin{equation}
 \gamma (t) = \gamma_S \, H_E \biggl( \theta - \frac{t}{T_I} \biggr)
 = \frac{2}{\beta} H_E \biggl( \theta - \frac{t}{T_I} \biggr) \, ,
 \end{equation}
where $H_E (x)$ is the Heaviside step function, using the half-maximum convention,
 which gives $1$ if  $x > 0$, gives $0.5$ if $x = 0$, and gives 0 if $x < 0$.
 $\theta$ is an adjusting value for $0 \leq \theta < 1$.
 The value $\theta$ is set to $0.5$ for example.
 Here, in this study, $t$ is the Monte Carlo time, and $T_I$ is the ending Monte Carlo time.
 The present QA schedule is made for the comparison between the system with no transverse field and the system with the smallest effective transverse field.
 The inverse temperature $\beta$ is set to a low but finite temperature.
 For $J (t)$ in Eq.~(\ref{eq:HJg}), using $J (t) = t / T_I$ for $0 \le t \le T_I$
 may be the simplest form as used in Ref.~\cite{FGGS}.
 Therefore, we investigate
\begin{equation}
 {\cal H}_1 (t) = - \frac{t}{T_I} \sum_{< i j >} \tau_{i j} \sigma^z_i \sigma^z_j
 - \frac{2}{\beta} H_E \biggl( 0.5 - \frac{t}{T_I} \biggr)  \sum_{i = 1}^N \sigma^x_i 
    \label{eq:Hamiltonian-1}
\end{equation}
 as the present-quantum-annealing schedule in this article.
 The Hamiltonian ${\cal H}_2 (t)$ 
 for a conventional-quantum-annealing schedule is given by \cite{FGGS}
\begin{equation}
 {\cal H}_2 (t) = - \frac{t}{T_I} \sum_{< i j >} \tau_{i j} \sigma^z_i \sigma^z_j
 - \biggl( 1 - \frac{t}{T_I} \biggr) \sum_{i = 1}^N \sigma^x_i 
 \, . \label{eq:Hamiltonian-2}
\end{equation}
 We also investigate ${\cal H}_2 (t)$ as a conventional-quantum-annealing schedule 
 in this article for comparison.
 As a classical-annealing schedule, we investigate
\begin{equation}
 {\cal H}_3 (t) = - \frac{t}{T_I} \sum_{< i j >} \tau_{i j} \sigma^z_i \sigma^z_j  \label{eq:Hamiltonian-3}
\end{equation}
 in this article for comparison.
 As for calculating annealing processes,
 this classical annealing schedule is mathematically the same with
 a classical annealing for $J = 1$, $\beta (t) = \beta t / T_I$ and $\gamma = 0$ in Eq.~(\ref{eq:H2}),
 thus we investigate ${\cal H}_3 (t)$.

In Refs.~\cite{KN, HRIT}, an another quantum-annealing schedule has also been applied.
 This QA schedule is that
 the transverse field $\gamma (t)$ is initially much larger than the couplings,
 $\gamma (0) \gg J (0) |\tau_{i j}|$, $J (t)$ is a constant for $t$, and,
 during QA, $\gamma (t)$ is slowly reduced to zero.
 For this annealing schedule, it is reported in Ref.~\cite{HRIT} that
 Monte Carlo simulations for the Ising spin glass model on the square lattice
 in the physically relevant continuous-imaginary-time limit
 do not show superiority of QA against CA.
 We do not investigate this schedule in this article
 because the results are already shown in Ref.~\cite{HRIT}.
 This schedule is the conventional schedule of the quantum annealing,
 and the schedule of ${\cal H}_2 (t)$ is the conventional schedule used in 
 a quantum computation by adiabatic evolution\cite{FGGS, FGGLLP}.
 The schedule of ${\cal H}_2 (t)$ has not been tested in the framework of the quantum annealing, thus we investigate the schedule of ${\cal H}_2 (t)$ in this article

 A method using a pulse of the transverse field is also proposed in Ref.~\cite{MSN},
 however the annealing schedule for this method is unclear in Ref.~\cite{MSN}.
 Thus we do not compare the present annealing schedule with this method in this article.

 We investigate the effectivities
 for ${\cal H}_1 (t)$, ${\cal H}_2 (t)$ and ${\cal H}_3 (t)$ in this article.

In this study, the exchange interaction is tuned instead of the temperature.
 Even if the temperature is tuned, the temperature is tuned from infinity to a low temperature since our aim is to simulate real-physical systems.
 It can be considered that the temperature does not reach absolute zero in real-physical systems.
 It can be reasonable that the final temperature of annealing is a low temperature.
 In this study, the exchange interaction changes instead of the temperature, thus, if the transverse field is imposed, the annealing becomes QA, and, if the transverse field is not imposed, the annealing becomes CA.

 By using Eqs.~(\ref{eq:a-13}) and (\ref{eq:c-2}) on condition of no kink
 and a fixed number for adding cuts,
 the transverse field is quantized as
\begin{equation}
 \gamma_Q (n) = \frac{n}{\beta} \, , \quad n = 2, 3, 4, \ldots \, , \label{eq:gQ}
\end{equation}
 in the Ising model for generating quantum effect.
 Even if a transverse field $\gamma_Q (n) $ is imposed,
 the quantum effect increases with increasing the number of kinks.
 However, to know that the transverse field is quantized
 as in Eq.~(\ref{eq:gQ}) can be convenient for adjustment of the quantum effect.
 $\gamma_Q (2)$ is the smallest effective field
 used in this study ($\gamma_Q (2) = \gamma_S$).
 If a transverse field $\gamma_Q$ for $n = 3$ is imposed,
 the generation of quantum effect for $n = 3$ is guaranteed at least.
 When a transverse field $\gamma_{\rm ex}$ is imposed for example,
 the quantum number $n_{\rm ex}$ is obtained as
 $n_{\rm ex} = \beta \gamma_{\rm ex}$.
 If $n_{\rm ex} < 2$,  $\gamma_{\rm ex}$
 is estimated as no effective for generating quantum effect.
  If $n_{\rm ex} \geq 2$, the generation of quantum effect
 for $n_{\rm ex}$ is guaranteed at least.
 Since very large $n$ disorders spin orders too much,
 searching proper $n$ for time $t$ is,
 of course, required for good performance of optimization.
 This quantization is a quantization for the number of cutting
 spin correlation per spin along the continuous-imaginary-time direction.
 Because the quantum effect by the transverse field depends on
 the number of the cuts for the correlation lengths
 along the continuous-imaginary-time direction,
 we focus on this quantization, and we utilize the fewest quantum number in this article.

\section{Simulation results} \label{sec:SimulationResults}

 We firstly tested our program source code for the two-spin-model case
 in order to fix software bugs, because
 static properties in the two-spin model are exactly calculated.
 We secondly tested our program source code for
 the one-dimensional ferromagnetic Ising spin chain with periodic boundary conditions
 under the transverse field ($\tau_{i j} = 1$ and $\gamma (t) = \gamma$) 
 in order to fix software bugs.
 The critical properties at $T_E = 0$ are identical
 to those of the classical two-dimensional Ising model and
  are well known exactly (the critical exponents
 are $\gamma_C = 1$, $\nu = 1$, $\beta = 1 / 8$
 and $z = 1$) \cite{MW}.
 We confirmed the values of
 the critical exponents $\nu$ and $\beta$ by using this algorithm
 and the finite-size-scaling method.
 We do not show the results of the two-spin-model case
 and the one-dimensional-ferromagnetic-Ising-spin-chain case,
 because the simulations were performed for fixing software bugs.
 Software bugs are fixed as above.

 Next, we performed simulations for the $\pm J$ Ising spin glass model on the square lattice with periodic boundary conditions.
 For each system size and each ending Monte Carlo time,
 $100$ realizations of $\tau_{i j}$ were investigated.
 For each realization, $10$ simulations were performed.
 Thus, $1000$ simulations were performed for each system size
 and each ending Monte Carlo time.
 $100$ Monte Carlo steps were used for preparing the initial states at each simulation.
 The inverse temperature $\beta$ was set to $\beta = 10$.
  The CA schedule,
  the conventional-quantum-annealing (CQA) schedule
 and  the present-quantum-annealing  (PQA) schedule were performed.
 All the instances used for CA, CQA and PQA are the same.

\begin{figure}[t]
\begin{center}
\includegraphics[width=0.98\linewidth]{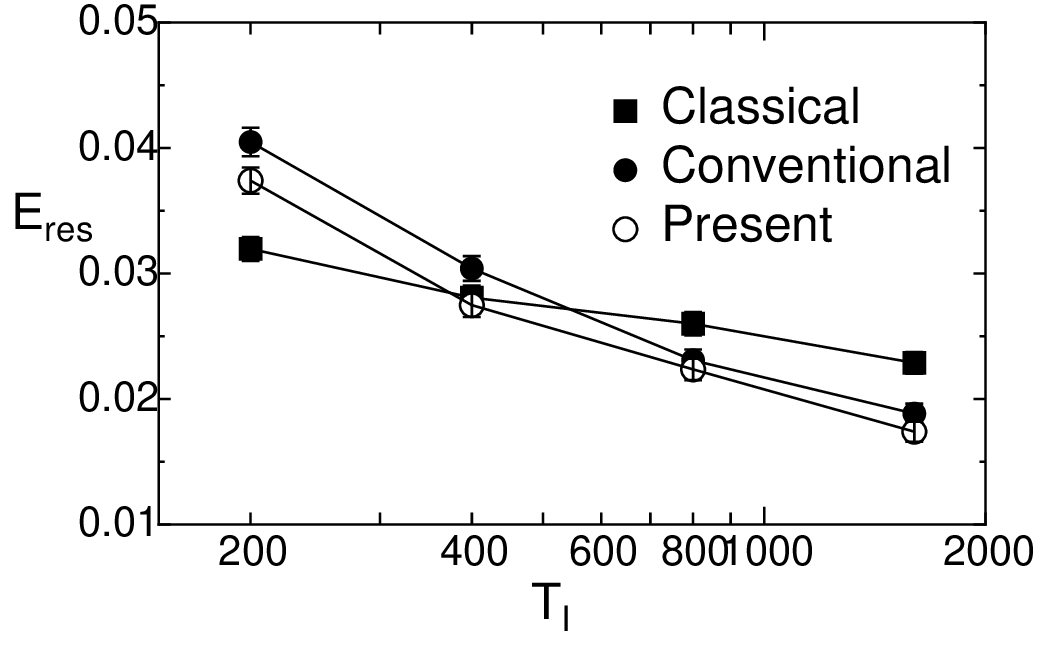}
\end{center}
\caption{
The relation between the ending Monte Carlo time $T_I$ and the residual energy $E_{\rm res}$.
The results of the $\pm J$ Ising spin glass model on the square lattice are shown.
The number of spins, $N$, is $100$, and the inverse temperature $\beta$ is $10$. 
The solid square represents the result for the classical-annealing schedule,
 the solid circle represents the result for the conventional-quantum-annealing  schedule,
 and  the open circle represents the result for the present-quantum-annealing  schedule.
\label{fig:Eres-TI}
}
\end{figure}
 Fig.~\ref{fig:Eres-TI} shows
 the relation between the ending Monte Carlo time $T_I$ and
 the residual energy $E_{\rm res}$ given in Eq.~(\ref{eq:Eres}).
The number of spins, $N$, is $100$. 
The solid square represents the result for the CA schedule,
 the solid circle represents the result for the CQA schedule,
 and  the open circle represents the result for the PQA schedule.
 At $T_I = 200$, the result of CA is better than the results of CQA and PQA.
 At $T_I = 1600$, the results of CQA and PQA are better than the result of CA.
 Therefore, by fast annealing, CA is seen to outperform QA, but,
 by slow annealing, QA is seen to outperform CA.
 At $T_I = 200$ and $T_I = 400$, the results of PQA are better than the results of CQA.

\begin{figure}[t]
\begin{center}
\includegraphics[width=0.98\linewidth]{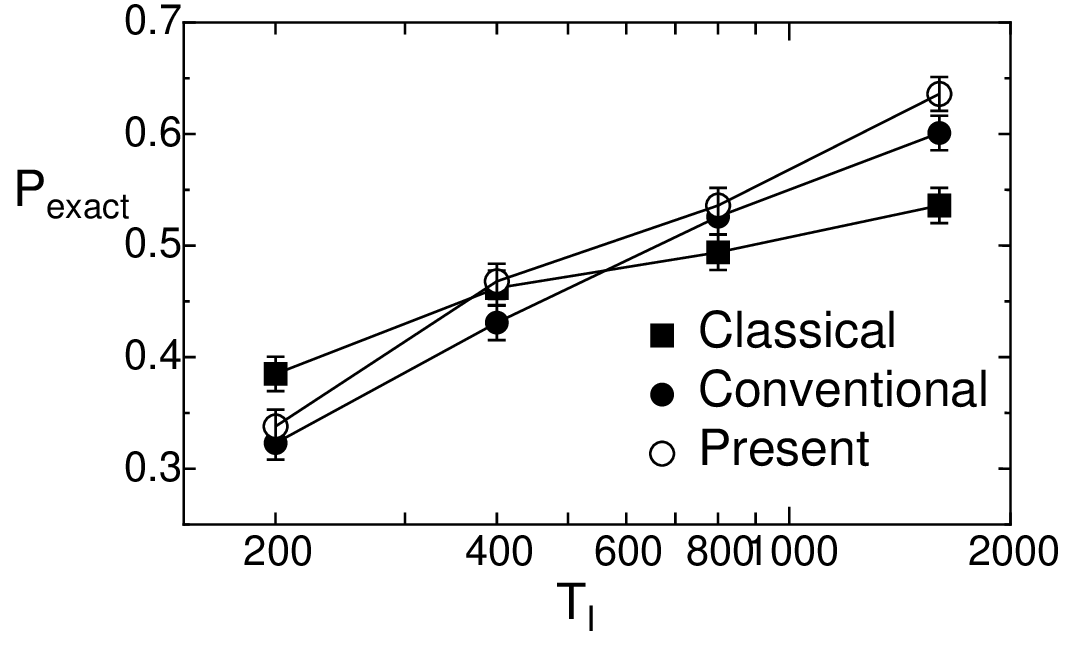}
\end{center}
\caption{
The relation between the ending Monte Carlo time $T_I$ and the accuracy $P_{\rm exact}$.
The results of the $\pm J$ Ising spin glass model on the square lattice are shown.
The number of spins, $N$, is $100$, and the inverse temperature $\beta$ is $10$. 
The solid square represents the result for the classical-annealing schedule,
 the solid circle represents the result for the conventional-quantum-annealing  schedule,
 and  the open circle represents the result for the present-quantum-annealing  schedule.
\label{fig:P-TI}
}
\end{figure}
 Fig.~\ref{fig:P-TI} shows
the relation between the ending Monte Carlo time $T_I$ and the accuracy $P_{\rm exact}$
 given in Eq.~(\ref{eq:Pexact}).
The number of spins, $N$, is $100$. 
The solid square represents the result for the CA schedule,
 the solid circle represents the result for the CQA schedule,
 and  the open circle represents the result for the PQA  schedule.
 At $T_I = 200$, the result of CA is better than the results of CQA and PQA.
 At $T_I = 1600$, the results of CQA and PQA are better than the result of CA.
 Therefore, by fast annealing, CA is seen to outperform QA, but,
 by slow annealing, QA is seen to outperform CA.
 At $T_I = 400$ and $T_I = 1600$, the results of PQA are better than the results of CQA.

 From Figs.~\ref{fig:Eres-TI} and \ref{fig:P-TI},
 by fast annealing, CA is seen to outperform QA at a number of spins, but,
 by slow annealing, QA is seen to outperform CA at the number of spins.
 From Figs.~\ref{fig:Eres-TI} and \ref{fig:P-TI},
 PQA is seen to outperform CQA.
 
\begin{figure}[t]
\begin{center}
\includegraphics[width=0.98\linewidth]{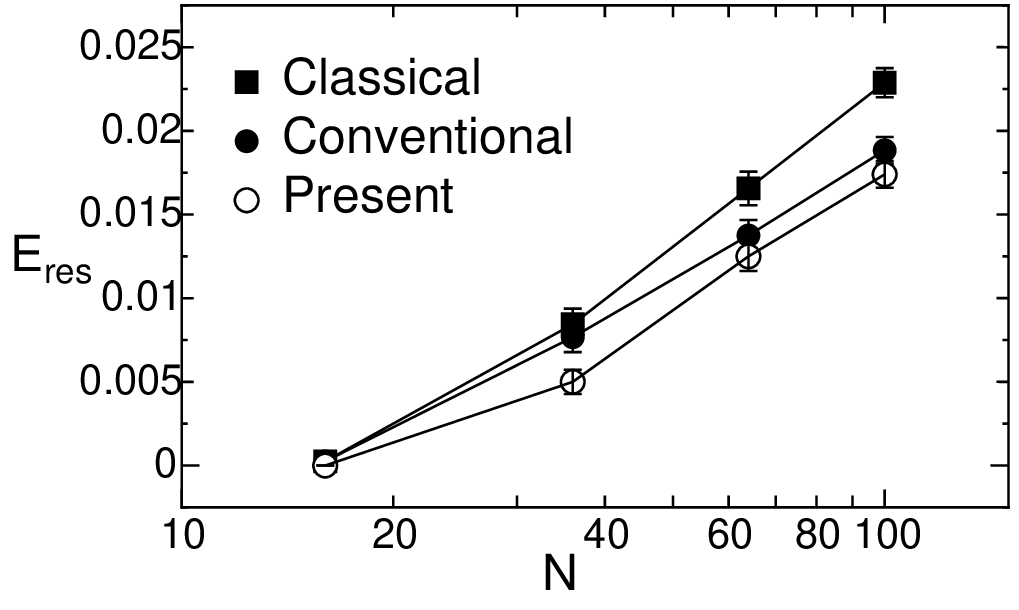}
\end{center}
\caption{
The relation between the number of spins, $N$, and the residual energy $E_{\rm res}$.
The results of the $\pm J$ Ising spin glass model on the square lattice are shown.
The ending Monte Carlo time $T_I$ is $1600$, and the inverse temperature $\beta$ is $10$. 
The solid square represents the result for the classical-annealing schedule,
 the solid circle represents the result for the conventional-quantum-annealing  schedule,
 and  the open circle represents the result for the present-quantum-annealing  schedule.
\label{fig:Eres-N}
}
\end{figure}
 Fig.~\ref{fig:Eres-N} shows
 the relation between the number of spins, $N$, and the residual energy $E_{\rm res}$
 given in Eq.~(\ref{eq:Eres}).
The ending Monte Carlo time $T_I$ is $1600$.
The solid square represents the result for the CA schedule,
 the solid circle represents the result for the CQA schedule,
 and  the open circle represents the result for the PQA schedule.
 At $N = 100$, the result of QA is better than the result of CA.
 QA is seen to outperform CA at large system sizes.
 At $N = 36$, the result of PQA is better than the result of CQA.
 We can see that the results of CQA and PQA are better
 than the result of CA for $N$ dependence.
 This shows the superiority of QA against CA.
 We could not see the superiority of PQA against CQA 
 for $N$ dependence.

\begin{figure}[t]
\begin{center}
\includegraphics[width=0.98\linewidth]{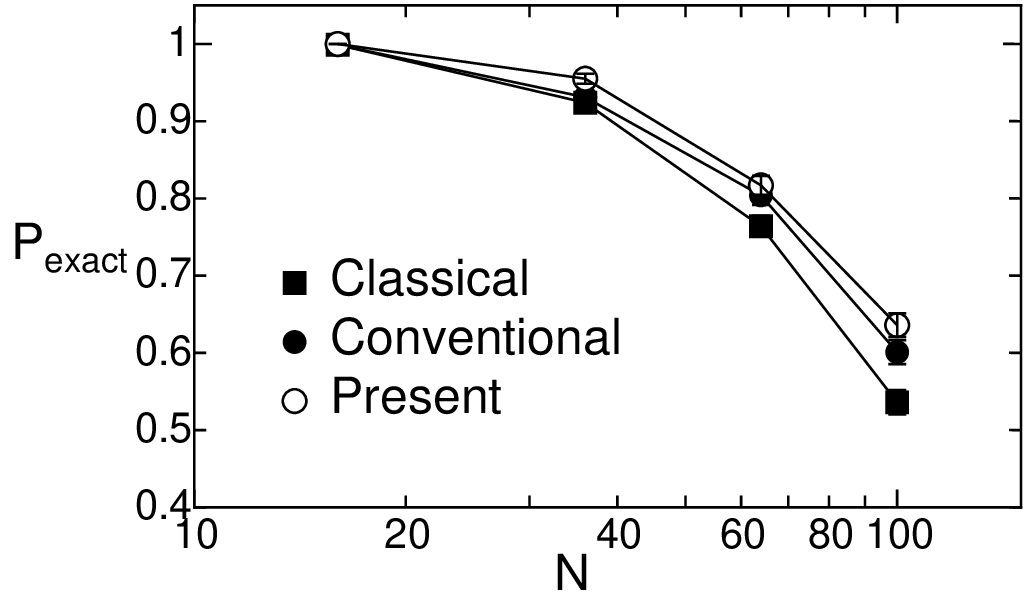}
\end{center}
\caption{
The relation between the number of spins, $N$, and the accuracy $P_{\rm exact}$.
The results of the $\pm J$ Ising spin glass model on the square lattice are shown.
The ending Monte Carlo time $T_I$ is $1600$, and the inverse temperature $\beta$ is $10$. 
The solid square represents the result for the classical-annealing schedule,
 the solid circle represents the result for the conventional-quantum-annealing  schedule,
 and the open circle represents the result for the present-quantum-annealing  schedule.
\label{fig:P-N}
}
\end{figure}
 Fig.~\ref{fig:P-N} shows
 the relation between the number of spins, $N$, and the accuracy $P_{\rm exact}$
 given in Eq.~(\ref{eq:Pexact}).
The ending Monte Carlo time $T_I$ is $1600$.
The solid square represents the result for the CA schedule,
 the solid circle represents the result for the CQA schedule,
 and the open circle represents the result for the PQA schedule.
 At $N = 100$, the result of QA is better than the result of CA.
 QA is seen to outperform CA at large system sizes.
 At $N = 36$ and $N = 100$,
 the results of PQA are better than the results of CQA.
 We can see that the results of CQA and PQA are better
 than the result of CA for $N$ dependence.
 This shows the superiority of QA against CA.
 We could not see the superiority of PQA against CQA 
 for $N$ dependence.

 From Figs.~\ref{fig:Eres-N} and \ref{fig:P-N},
  QA is seen to outperform CA for the dependences of the number of spins.
 From Figs.~\ref{fig:Eres-N} and \ref{fig:P-N},
 PQA is seen to outperform CQA.

\section{Concluding remarks} \label{sec:ConcludingRemarks}

 We performed Monte Carlo simulations
 for the spin glass model on the square lattice.
 As the simulation results,
 QA was seen to outperform CA at a number of spins
 when the annealing time was sufficiently spent.
 As the simulation results,
 QA was seen to outperform CA
 for the dependences of the number of spins
 when the annealing time was sufficiently spent.
 Therefore, our conclusion is that there is a superiority of QA against CA
 for the spin glass model by using Monte Carlo simulations.

 QA was seen to outperform CA
 for the system-size dependences.
 However, we were not able to numerically estimate
 whether the dependences are power dependences or
 exponential dependences,
 because the dependences were not clear enough.
 Although the numerical estimations for the system-size dependences
 would be hard tasks,
 the estimations are tasks for the future.

Conventionally, operating the temperature $T_E$
 with a fixed $J (t) (= J)$ and no $\gamma (t)$ term of Eq.~(\ref{eq:HJg})
 is called CA, however, in this article, we called that
 operating $J (t)$ with a fixed $T_E$ and no $\gamma (t)$ term of Eq.~(\ref{eq:HJg}) is CA.
 There are two reasons. One is that, as for calculating annealing processes,
 operating $T_E$ with a fixed $J$ and no $\gamma (t)$ term
 is mathematically the same with
 operating $J (t)$ with a fixed $T_E$ and no $\gamma (t)$ term.
 One is that operating $J (t)$ with a fixed $T_E$ and no $\gamma (t)$ term
 is more practical in relation to the D-Wave chip\cite{JAGetc}.
 These may be trivial, but we consider that these are also significant
 for understanding quantum annealing computing.

In this article, results at a low temperature are only shown.
 There is a problem of whether a temperature, that changes the superiority QA over CA, exists or not.
 However, investigating each optimized temperatures for each annealings can be more important.
 These are tasks for the future

The size of the transverse field used in the conventional QA schedule
 is larger than the size of that used in the present QA schedule,
 however, as the simulation results, 
 the present QA schedule was more effective than the conventional QA schedule.
 The reason may be that
 the relaxation time for disorders by large transverse field is required,
 although the tunneling effect by the transverse field
 helps to make the systems escape
 from local minima of the free energy.
 Therefore, this means that
 a thorough investigation for setting of QA schedule
 is needed.
 This can be an important task for the future.

Very recently, it is reported in Ref.~\cite{HBHKMT} that, by fast annealing,
 QA is seen to outperform CA for their benchmark spin-glass problems.
 On the other hand, our results showed that, by slow annealing,
 QA is seen to outperform CA for our benchmark spin-glass problems.
  Since the used annealing schedules and the used models are different,
 we can not say anything for the differences between the results, but
 they have also proposed a way of comparing QA with CA.
 Their schedule uses observables, on the other hand,
 our schedule uses a smallest effective field.
 Their schedule can be a better schedule,
 however their schedule depends on models.
 On the other hand, our schedule does not depend on models.
 For this difference between the QA schedules,
 we believe that the present study is also significant.
  The detailed comparison is a task for the future.

 Other QA schedules using the smallest effective field derived in this article
 may also be worth considering.
 The present QA schedule is made based on the concept of perturbation theory.
 Other QA schedules using the smallest effective field
 based on the same concept can also be considered,
 and some schedules among them may be better for performance of optimization.

 A set of the computer program source codes, used in this study,
 is available on payment basis \cite{TQC}.

\section*{Acknowledgments} \label{Acknowledgments}

The spin glass server \cite{SGServer} was
used to obtain the ground-state energies for realizations.
We thank the persons concerned with the spin glass server.


\begin{thebibliography}{599}
\bibitem{RCC}
 P.~Ray, B.~K.~Chakrabarti and A.~Chakrabarti, {\it Phys.\ Rev.\ B } \textbf{39} (1989)
11828.
\bibitem{FGSSD}
 A.~B.~Finnila, M.~A.~Gomez, C.~Sebenik, C.~Stenson and J.~D.~Doll,
 {\it Chem.\ Phys.\ Lett.\ } \textbf{219} (1994) 343.
\bibitem{KN}
 T.~Kadowaki and H.~Nishimori,
 {\it Phys.\ Rev.\ E } \textbf{58} (1998) 5355.
\bibitem{ST}
 G.~Santoro and E.~Tosatti,
 {\it J.\ Phys.\ A:\ Math.\ Gen.\ } \textbf{39} (2006) R393.
\bibitem{DC}
 A.~Das and B.~K.~Chakrabarti,
 {\it Rev.\ Mod.\ Phys.\ } \textbf{80} (2008) 1061. 
\bibitem{TTC}
 S.~Tanaka, R.~Tamura and B.~K.~Chakrabarti, {\it Quantum Spin Glasses, Annealing and Computation} (Cambridge University Press, Cambridge and Delhi, 2017).
\bibitem{FGGS}
 E.~Farhi, J.~Goldstone, S.~Gutmann and M.~Sipser, arXiv:quant-ph/0001106.
\bibitem{FGGLLP}
 E.~Farhi, J.~Goldstone, S.~Gutmann, J.~Lapan, A.~Lundgren and D.~Preda,
 {\it Science} \textbf{292} (2001) 472.
\bibitem{GG}
 S.~Geman and D.~Geman, {\it IEEE Trans. Pattern Anal. Mach. Intell.} \textbf{6} (1984)
721.
\bibitem{JAGetc}
 M.~W.~Johnson, M.~H.~S.~Amin, S.~Gildert, T.~Lanting, F.~Hamze,
 N.~Dickson, R.~Harris, A.~J.~Berkley, J.~Johansson, P.~Bunyk, E.~M.~Chapple,
 C.~Enderud, J.~P. Hilton, K.~Karimi, E.~Ladizinsky, N.~Ladizinsky,
 T.~Oh, I.~Perminov, C.~Rich, M.~C.~Thom, E.~Tolkacheva, C.~J.~S.~Truncik,
 S.~Uchaikin, J.~Wang, B.~Wilso and G.~Rose, {\it Nature} \textbf{473} (2011) 194.
\bibitem{HRIT}
 B.~Heim, T.~F.~R\o nnow, S.~V.~Isakov and M.~Troyer, {\it Science} \textbf{348} (2015) 215.
\bibitem{O}
 A.~T.~Ogielski, {\it Phys.\ Rev.\ B\ } \textbf{32} (1985) 7384.
\bibitem{D}
 B.~Derrida, {\it Phys.\ Rep.\ } \textbf{184} (1989) 207.
\bibitem{NI}
 T.~Nakamura and Y.~Ito, {\it J.\ Phys.\ Soc.\ Jpn.\ } \textbf{72} (2003) 2405.
\bibitem{RK}
 H.~Rieger and N.~Kawashima, {\it Euro.\ Phys.\ J.\ B\ } 9  (1999) 233.
\bibitem{MSN}
 S.~Morita, S.~Suzuki and T.~Nakamura,
 {\it Phys.\ Rev.\ E\ } \textbf{79} (2009) 065701.
\bibitem{EA}
 S.~F.~Edwards and P.~W.~Anderson, {\it J.\ Phys.\ F\ } \textbf{5} (1975) 965.
\bibitem{MPV}
 M.~M\'{e}zard, G.~Parisi and M.~A.~Virasoro, {\it Spin Glass Theory and Beyond}
 (World Scientific, Singapore, 1987).
\bibitem{N}
 H.~Nishimori, {\it Statistical Physics of Spin Glasses and Information Processing: An Introduction} (Oxford University Press, Oxford, UK, 2001).
\bibitem{KR}
 N.~Kawashima and H.~Rieger,
 Recent progress in spin glasses, in {\it Frustrated Spin Systems}, ed. H.~T.~Diep (World Scientific, Singapore, 2004).
\bibitem{S}
 M.~Suzuki, {\it Prog.\ Theor.\ Phys.\ } \textbf{56} (1976) 1454.
\bibitem{KG}
 N.~Kawashima and J.~E.~Gubernatis,
  {\it Phys.\ Rev.\ E\ } \textbf{51} (1995) 1547.
\bibitem{MW}
 B.~M.~McCoy and T.~T.~Wu, {\it The two-dimensional Ising model}
 (Harvard University Press, Cambridge, Massachusetts, 1973).
\bibitem{HBHKMT}
 D.~Herr, E.~Brown, B.~Heim, M.~K\"onz, G.~Mazzola and M.~Troyer, arXiv:1705.00420.
\bibitem{TQC}
 http://www.to-qc.com/services/.
\bibitem{SGServer}
 http://www.informatik.uni-koeln.de/spinglass/.
\end{thebibliography}
\end{document}